# Investigation of high voltage discharges in low pressure gases through large ceramic superconducting electrodes

Evgeny Podkletnov[1], Giovanni Modanese[2]

[1]Moscow Chemical Scientific Research Center
113452 Moscow – Russia
E-mail: epodkletnov@hotmail.com

[2]*University of Bolzano*
Logistics and Production Engineering
*Via Sernesi 1, 39100 Bolzano - Italy*
E-mail: giovanni.modanese@unibz.it

**Abstract:** A device has been built and tested, in which a ceramic superconducting cathode and a copper anode cause electrical discharges in low pressure gases, at temperatures between 50 and 70 *K*. The electrodes are connected to a capacitors array charged up to 2000 *kV*; peak currents are of the order of $10^4$ *A*. The cathode has the diameter of 10 *cm* and is fabricated by OCMTG technology. In discharges at voltage above 500 *kV* two new phenomena were observed, probably related to each other. First, the discharge does not look like a spark, but is a flat, glowing discharge, which originates from the whole surface of the superconducting electrode. Furthermore, a radiation pulse is emitted at the discharge, which propagates orthogonally to the cathode, towards the anode and beyond it, in a collimated beam, apparently without any attenuation. The radiation pulse carries an energy of $10^{-3}$ *J* at least. The features and the nature of this radiation have been investigated by several means, still it was not possible to identify it; we can only exclude that it is electromagnetic radiation or any other radiation with energy-momentum relationship *E=cp*.

## 1. Introduction

Recent developments of the OCMTG technology (oxygen controlled melt texture growth) made possible to manufacture large elements of high quality ceramic superconductors with pre-definite crystal structure. This fabrication technique requires careful control of all the stages of the sintering and crystal growth process, but offers in return several different opportunities, either for construction of novel electrical-magnetic devices, or for improvements in the performance of conventional devices. For instance, YBCO coatings and bulk YBCO elements have been employed in several applications to passive bearings [1] and in MHD generators ([2] and ref.s). The apparatus described in this paper represents the evolution of a series of Van de Graaf generators modified by the application of a YBCO coating to one of the (spherical or toroidal) electrodes. These were used to organize high voltage discharges in rarefied gases, varying the gas composition and pressure, as well as temperature, electrodes distance, containment magnetic field etc.

The motivation for this research originated in part from some works by Tonouchi et al. [3,4]. They investigated the behavior of ceramic superconducting films subjected to high frequency perturbations which led to non-equilibrium, dissipative situations. In certain conditions, emission of a peculiar electromagnetic radiation was also observed, whose features are determined by those of microscopic antennas carved in the ceramic films. In our apparatus an emission from the superconducting electrode was observed, too. It does not appear, however, to be of electromagnetic nature. Our electrodes being made of well oriented bulk crystals, we believe that the features of this radiation are essentially determined by the lattice parameters of the material.

The experimental apparatus is described in Section 2. Section 3 contains the results of the measurements and Section 4 a discussion of these results. In Section 5 a hint to a possible theoretical explanation is given. Section 6 comprises our conclusions and outlook.

## 2. Experimental

### 2.1 General description of the installation

The experimental set-up consists of a discharge chamber and an electric block with a Marx high voltage generator that is able to create a voltage up to 2000 *kV*. The scheme of the discharge chamber is presented in Fig.1 (top view). The chamber has the form of a cylinder with the diameter of about 1 *m* and the length of 1.5 *m* and is made of 3 sections of tubes of silica quartz, thus allowing observation of the shape, the trajectory and the colour of the discharge. The discharge is organized between the superconducting ceramic emitter and a target electrode. The superconducting emitter has the shape of a disk with round corners and consists of two layers (see Section 2.2). The non-superconducting part of the emitter is fixed to a copper cryostat using metal indium or Wood's metal, the superconducting part of the emitter faces the target electrode. The target is a copper disk with the diameter of 100 *mm* and the thickness of 15 *mm*. The design permits the creation of high vacuum inside the chamber or to fill the whole volume with any gas. The distance between the electrodes can vary from 15 to 40 *cm* in order to find the optimum length for each type of the emitter. The chamber has a connecting section with flanges which allows easy changing of the emitter.

The discharge can be concentrated on a smaller target area using a large solenoid with the diameter of 1.05 *m* that is wound around the chamber using copper wire with the diameter of 5 *mm*. The length of this coil is about 30 *cm* and there are several axial layers, so that the thickness of the wounding is about 10 *cm*. The magnetic flux density is 0.9 *T*. A small solenoid is also wound around the emitter (Fig. 1) so that some magnetic flux can be trapped inside the emitter when it is cooled down below the critical temperature.

The refrigeration system for the superconducting emitter provides a sufficient amount of liquid nitrogen or liquid helium for the long-term operation and the losses of gas due to evaporation are minimized because of the high vacuum inside the chamber. The superconducting electrode during the discharge is kept at the temperature between 40 and 80 *K*. The temperature of the superconductor is measured using a standard thermocouple for low temperature measurements.

A photodiode is placed on the transparent wall of the chamber and is connected to an oscilloscope, in order to provide information on the light parameters of the discharge.

A precise measurement of the voltage of the discharge is achieved using a capacitive sensor that is connected to an oscilloscope with a memory option as shown in the upper part of Fig.1. Electrical current measurements are carried out using a Rogowski belt, which is a single loop of a coaxial cable placed around the target electrode and connected to the oscilloscope.

Our high voltage Marx generator (as shown in Fig. 2) consists of twenty capacitors (25 *nF* each) connected in parallel and charged to a voltage up to 50-100 *kV* using a high voltage transformer and a diode bridge. The capacitors are separated by resistive elements of about 100 *k*Ω. The scheme allows charging of the capacitors up to the needed voltage and then to change the connection from a parallel to a serial one. The required voltage is achieved by changing the length of the air gap between the contact spheres C and D. A syncro pulse is then sent to the contacts C and D which causes an overall discharge and serial connection of the capacitors and provides a powerful impulse of up to 2000 $kV$ which is sent to the discharge chamber.

In order to protect the environment and the computer network from static electricity and powerful electromagnetic pulses, the chamber is shielded by a Faraday cage with the cell dimensions of 2×2 $cm$ and a rubber-plastic film material absorbing ultra high frequency (UHF) radiation.

## 2.2 Superconducting emitter, fabrication methods

The superconducting emitter has the shape of a disk with the diameter of 80-120 $mm$ and the thickness of 7-15 $mm$. This disk consists of two layers: a superconducting layer with chemical composition $YBa_2Cu_3O_{7-y}$ (containing small amounts of *Ce* and *Ag*) and a layer with chemical composition $Y_{1-x}Re_xBa_2Cu_3O_{7-y}$, where *Re* represents *Ce*, *Pr*, *Sm*, *Pm*, *Tb* or other rare earth elements. This second layer is not superconducting in the temperature range of interest. The materials of both layers were synthesized using a solid state reaction under low oxygen pressure (stage 1), then the powder was subjected to a melt texture growth (MTG) procedure (stage 2). Dense material after MTG was crushed, ground and put through sieves in order to separate the particles with the required size. A bi-layered disk was prepared by powder compaction in a stainless steel die and sintering using seeded oxygen controlled melt texture growth (OCMTG) (stage 3). After mechanical treatment the ceramic emitter was attached to the surface of the cooling tank in the discharge chamber using an Indium based alloy. The fabrication stages are described below.

**Stage 1 -** Micron-size powders of $Y_2O_3$ and *CuO*, $BaCO_3$ were mixed in alcohol for 2 hours, then dried and put in zirconia boats in a tube furnace for heat treatment. The mixture of powders was heated to 830° *C* and kept at this temperature for 8 hours at oxygen partial pressure of $2.7 \cdot 10^2$ *Pa* (or 2-4 *mBar*) according to [5,6]. The material of the normal conducting layer was sintered in a similar way.

**Stage 2 -** Micron-size powder of $YBa_2Cu_3O_x$ was pressed into pellets using a metal die and low pressure. The pellets were heated in air to 1050° *C* (100° *C* per hour), then cooled to 1010° *C* (10° *C* per hour), then cooled to 960° *C* (2° *C* per hour), then cooled to room temperature (100° *C* per hour) according to a standard MTG technique [7,8]. The quantity of 211 phase during heating was considerably reduced and the temperature was changed correspondingly. $ReBa_2Cu_3O_{7-x}$ was also prepared using MTG, but the temperature was slightly changed according to the properties of the corresponding rare earth oxide.

**Stage 3 -** Bulk material after MTG processing was crushed and ground in a ball mill. The particles with the size less than 30 $\mu m$ were used for both layers of the ceramic disk. The particles were mixed with polyvinyl alcohol binder. The material of the first layer was put into a die, flattened and then the material of the second layer was placed over it. The disk was formed using a pressure of 50 *MPa*. The single crystal seeds of *Sm*123 (about 1 $mm^3$) were placed on the surface of the bi-layered disk so that the distance between them was about 15 mm and the disk was subjected to a OCMTG [9,10] treatment in 1% oxygen atmosphere. The growth kinetics of YBCO superconductor were controlled during isothermal melt texturing. A modified melt texturing process was applied, where instead of slow cooling following melting, isothermal hold was employed in the temperature range where the growth is isotropic. By this modification, the time required to texture the disk was reduced to 7 hours which is about 10 times faster than a typical slow cooling melt texturing process. The crystallization depth was controlled by applying the corresponding temperature and time parameters. Cubic *Sm*123 seeds were obtained using the nucleation and growth procedure as described in [11,12]. A thin layer of the material was removed from the top surface of the disk to a depth of 0.3 *mm* and the edges of the upper surface were rounded using diamond tools.

X-ray crystal lattice parameters, transition temperature, electrical conductivity and critical current density were measured for both layers of various emitters using standard techniques.

**2.3 Organization of the discharge and measurements of the effect**

The discharge chamber is evacuated to 1.0 *Pa* using first a rotary pump and then a cryogenic pump. When this level of vacuum is reached, liquid nitrogen is pumped into a tank inside the chamber that contacts the superconducting emitter. Simultaneously a current is sent to the solenoid that is wound around the emitter, in order to create a magnetic flux inside the superconducting ceramic disk. When the temperature of the disk falls below the transition temperature (usually 90 *K*) the solenoid is switched off. The experiment can be carried out at liquid nitrogen temperatures or at liquid helium temperatures. If lower temperatures are required, the tank is filled with liquid helium and in that case the temperature of the emitter reaches 40-50 *K*.

The high voltage pulse generator is switched on and the capacitors are charged to the required voltage. It takes about 120 *s* to charge the capacitors. A syncro pulse is sent to a pair of small metal spheres marked as C and D in Fig. 2. A discharge with voltage up to 2000 *kV* occurs between the emitter and the target. Half a second before the discharge, a short pulse of direct current is sent for 1 *s* to the big solenoid that is wound around the chamber, in order to concentrate the discharge on the target. This pulse lasts for only 1 *s* not to cause the overheating of the big solenoid.

The anomalous radiation emitted (see Section 3) is measured along the projection of the axis line which connects the center of the emitter with the center of the target. Laser pointers were used to define the projection of the axis line and impulse sensitive devices were situated at the distance of 6 *m* and 150 *m* from the installation (in another building across the area).

Normal pendulums were used to measure the radiation pulses coming from the emitter. The pendulums consisted of spheres of different materials hanging on cotton strings inside glass cylinders under vacuum. One end of the string was fixed to the upper cap of the cylinder, the other one was connected to a sphere. The spheres had typically a diameter from 10 to 25 *mm* and had a small pointer in the bottom part. A ruler was placed in the bottom part of the cylinder, 2 *mm* lower than the pointer. The length of the string was typically 80 *cm*. The deflection was observed visually using a ruler inside the cylinder; in fact, there are significant variations in the amplitude of the impulse for repeated discharges (Section 3), so the visual observation does not substantially affect precision.

Various materials were used as spheres in the pendulum: metal, glass, ceramics, wood, rubber, plastic. The tests were carried out when the installation was covered with a Faraday cage and UHF radiation absorbing material and also without them. The installation was separated from the impulse measuring devices situated 6 *m* away by a 30 *cm* thick brick wall and a 25 *mm* thick list of steel with the dimensions 1×1.2 *m*. The measuring systems that were situated 150 *m* away were additionally shielded by brick walls of 80 *cm* total thickness. In the latter case, the laser pointers' rays made a right angles detour through nearby windows.

In order to define some other characteristics of the impulse - in particular its frequency spectrum - a condenser microphone was placed along the impact line just after the glass cylinders. The microphone was placed in a plastic spherical container filled with porous rubber. The microphone was first oriented with a membrane facing the direction of the discharge, then it was turned 22.5 degrees to the left, then 45 degrees to the left, then 67.5 degrees and finally 90 degrees. Several discharges were recorded in all these positions at equal discharge voltage. The coincidence with the discharge was only observed visually.

In order to assess the effect of the anomalous radiation on light, some preliminary measurements with a laser beam were done. The interaction turned out to be small, so we acted as to maximize the interaction region: In the configuration with discharge chamber and mechanical detectors placed 150 *m* apart, the laser beam was brought to cross the anomalous radiation beam with an angle of approx. 0.1°, giving an interaction region ca. 57 *m* long. We mainly used a ruby laser with wavelength 694 *nm*, power from 0.5 to 10 *W* and spot diameter about 4 *mm*. A blue laser with wavelength 473 *nm*, power of 0.45 *W* and spot diameter of 3 *mm* was also employed. The intensity of the laser light was monitored through a Femtosecond Optically Gated Fluorescence Kinetic Measurement System FOG100.

## 3. Results

The discharge at room temperature in the voltage range from 100 *kV* to 2000 *kV* was similar to a discharge between metal electrodes and consisted of a single spark. When the superconductor was cooled down below the transition temperature, the shape of the discharge changed in such a way that it did not form a direct spark between the two electrodes, but the sparks appeared from many points on the superconducting emitter and then moved to the target (positive) electrode. When the voltage was increased to 500 *kV* the front of the moving discharge became flat with the diameter corresponding to that of the emitter. This flat glowing discharge separated from the emitter and moved to the target electrode with great speed. For maximum distance between the emitter and the target (about 1 *m*) it is possible to see the flat glowing sparkle that jumps from the emitter to the target. When the distance is reduced to 0.25 *m* the time of the discharge as defined by the photo diode is between $10^{-5}$ and $10^{-4}$ *s*. The peak value of the current at the discharge for the maximum voltage (2000 *kV*) is of the order of $10^4$ A.

Given the low pressure and the high applied voltage, emission of X-rays from the metallic electrode cannot be excluded, but the short duration of the discharge makes their detection difficult. Use of a Geiger counter and of X-rays sensitive photographic plates did not yield any clear signature of X-rays.

It was found that high voltage discharges organized through the superconducting emitter kept at the temperature of 50-70 *K* were accompanied by a very short pulse of radiation coming from the superconductor and propagating along the axis line connecting the center of the emitter and the center of the target electrode in the same direction as the discharge. The radiation appeared to penetrate through different bodies without any noticeable loss of beam strength. It acted on small interposed mobile objects like a repulsive force field, with a force proportional to the mass of the objects.

The dependence of the effect on the temperature (in the range between 50 and 70 *K*) and on the duration of the high voltage pulse was not significant.

In order to investigate the interaction of this radiation with various materials, several tests were carried out, with pendulums and microphones, as described in the experimental part. The correlation between the discharge voltage and the corresponding horizontal deflection of the pendulum as measured for two different emitters is shown in Fig. 3. Each value of Δ*l* in the figure represents the average figure calculated from 12 discharges. A rubber sphere with the weight of 18.5 *g* was used as material of the pendulum.

Both emitters, N. 1 and N. 2, were manufactured using the same OCMTG technology, but the thickness of the superconducting layer was equal to 4 *mm* for the emitter N. 1 and 8 *mm* for the emitter N. 2. Emitter N. 2 could be magnetized to a much higher value. The thickness of the normal conducting layer has a smaller influence on the force of the impulse, but for better results the thickness should be bigger than 5 *mm*. As a check, conventional emitter materials were also used, such as copper, aluminium, steel, chromites. They gave no unusual effects.

It was found that the force of the impact on pendulums made of different materials does not depend on the material but is only proportional to the mass of the sample. Pendulums of different mass demonstrated equal deflection at constant voltage. This was proved by a large number of measurements using spherical samples of different mass and diameter. The range of the employed test masses was between 10 and 50 *g*. The pendulum bobs did not show any signs of heating after repeated pulses. It was also found that there exist certain deviations in the force of the impulse within the area of the projection of the emitter. These deviations (up to 12-15% max) were found to be connected with the inhomogeneities of the emitter material and various imperfections of the crystal structure of the ceramic superconductor, and with the thickness of the interface between the superconducting and normal conducting layers.

Measurements of the impulse taken at close distance (3-6 *m*) from the installation and at the distance of 150 *m* gave identical results, within the experimental errors. As the points of measurements were separated by a thick brick wall and by air, it is possible that the losses in the media were negligible. The force beam does not appear to diverge and its borders are clear-cut. The beam size was measured up to ca. 5 *mm* by means of special boards that receive an imprint under the action of small pressure, similar to those commercially available (for instance from Sensor Products, East Hanover).

We did not notice any recoil on the apparatus after the discharges, however precise measurements with strain gauges or similar were not done.

The bi-layered emitters used in this experiment had a structure typical for multiple-domain levitators with well crystallized and oriented grains of the superconducting layer. The superconducting layer consisted of *$YBa_2Cu_3O_{7-y}$* orthorhombic superconductor with lattice parameters *a* = 3.89 Å, *b* = 11.69 Å, *c* = 3.82 Å. The addition of small amounts of *$CeO_2$* led to an improvement in the magnetic flux pinning properties of the Y123 compound. The superconducting layer had a maximum trapped field of 0.5 *T* at 77 *K* and a critical current density in excess of $5 \cdot 10^4$ *$A/cm^2$*. The transition temperature varied from 87 to 90 *K* with a transition width of about 2 degrees. The normal conducting layer had crystal lattice parameters close to those of the superconductor: *a* = 3.88 Å, *b* = 11.79 Å, *c* = 3.82 Å. Both layers demonstrated high electrical conductivity (over 1.5 *$Sm^{-1}$*) at room temperature and the *$Y_{1-x}Re_xBa_2Cu3O_{7-y}$* layer was a normal conductor above 20 *K*.

The role of the normal layer is not entirely clear yet. If it is absent, the discharges tend to be less regular and the anomalous radiation is much weaker. This might happen because the layer prevents a direct metal-superconductor junction, or because its ohmic resistance controls the discharge current.

The presence of trapped magnetic flux in the emitter was found to lead to an increase in the impulse strength of approximately the 25%. Therefore, at recent stages of the experiment the solenoid was replaced by a powerful permanent NdFeB magnet (the maximum energy product value of 50 *MGOe*) with a diameter corresponding to the diameter of the emitter and a thickness of 20 mm. This disk-shaped magnet was attached with one surface to the cooling tank and with another surface to the ceramic emitter.

The response recorded by the microphone has the typical behavior of an ideal pulse filtered by the impulse response of a physical low pass system with a bandwidth of about 16 *kHz*, attributed to the microphone (Fig. 4). In spite of the filtering, the relative energy of the pulses can be measured as a function of the angle of the normal to the diaphragm with respect to the axis of propagation of the force. Relative pulse amplitude with energies averaged over four pulses per angle are shown in Fig. 5 and are in agreement with a possible manifestation of a vector force acting directly on the membrane. No signal has been detected outside the impact region.

If the radiation beam propagates in air, some energy should be depleted from it as it propagates. If air molecules are accelerated exactly like the pendulums, at standard temperature and pressure the energy lost to air from the beam should be on the order of $10^{-3}$ *J/m*. The velocity shear between the air in the path of the impulse and surrounding atmosphere should, in principal, lead to noticeable air turbulence after the pulse. While extensive studies of the behavior of the air in the radiation beam have not been conducted, observations of the air in the beam path with smoke show that only brief forward and back movement of the particles occurs. There is no significant airflow, as the impulse is very short, and neither turbulence nor vortex phenomena are observed.

The interaction of the laser beam with the anomalous radiation in a region having the length of ca. 57 *m* caused the intensity of the initial laser spot to decrease by 7-10% during the discharge and then return quickly to baseline: our sensor indicates a rise time of $10^{-7}$ *s* or less.

We would like to stress that measurements with the laser beam should still be regarded as preliminary, because their results could in principle be affected by air density variations brought about by the anomalous radiation. The resulting refraction could confuse the true "primary" interaction of the radiation with light. Our results seem to exclude this *a posteriori*, because if air density variations really caused a spurious deviation or diffusion of the laser beam, its intensity could not be restored so quickly after the discharge; we expect on general grounds that density variations in air at room temperature and pressure would persist for at least $10^{-4}$-$10^{-5}$ *s*.

It would be desirable in the future to do measurements in vacuum; this is hardly compatible, however, with the need for a long interaction region in order to obtain a detectable signal. On the other hand, a short interaction region will be necessary if we want to use a laser beam and fast opto-electronic switches to measure the beam velocity (see the Conclusions Section).

## 4. Discussion

### 4.1 Phenomenological analysis

The features of discharges obtained with large superconducting emitters above 500 *kV*, as described in Section 3, appear to be unique and very interesting. The occurrence of the observed "flat" discharge could be explained as follows. The fabrication method described in Section 2, in particular when OCTMG is employed, gives the emitter a special crystal structure, with high-conductivity *ab* planes mostly parallel to the surface. (We recall that the penetration depth λ of YBCO is larger in the *c* direction than in the *ab* direction; typically one has $\lambda_{ab}$~135 *nm* and $\lambda_c$~890 *nm* [13].) This makes the surface of the emitter, to a high accuracy, a surface of constant electric potential. In usual spark discharges in a neutral gaseous medium the dielectric is first "broken" at some point near an electrode where the electric field is strong enough to generate cascade ionization. With our large superconducting electrode, the field at the electrode's surface is extremely uniform and a flat discharge starts only when its strength exceeds everywhere the value necessary for ionization.

At the discharge, the supercurrent flows through the emitter in the *c* crystal direction, except at grain boundaries and defect points where the material is non-isotropic. It is known that conduction in the *c* direction takes place only through a sort of tunneling between the *ab* planes. This does not affect the overall conduction rate, however, which is essentially limited by the normal layer of the emitter.

The discharge current crosses two SN junctions, the first one between the normal and superconducting part of the emitter, and the second one between the external surface of the emitter and the ionized gas. From the theoretical point of view, these junctions can be described in principle by Ginzburg-Landau models without specific reference to the microscopic theory. A typical time scale for the time-dependent Ginzburg-Landau equation in YBCO is of the order of $10^{-8}$ *s* [14], much smaller than the discharge time, and so a quasi-stationary approximation would be adequate in our case.

Let us estimate the current density *j* and the average velocity *v* of the superconducting charge carriers during the discharge. Taking an emitter area of 100 $cm^2$ and $I=10^4$ *A*, one finds $j=10^8$ $A/m^2$. For a carriers density of ~ $3 \cdot 10^{27}$ $m^{-3}$ [13] we have $v \sim 10^{-3}$ *m/s*. This implies that in the discharge time the superconducting charge carriers move on the average by $\Delta x \sim 10^{-7}$ *m*, going through ~1000 *ab* crystal planes.

The electrostatic energy stored in the capacitors of the Marx generator at the maximum voltage of 2000 *kV* is of the order of $10^6$ *J*. It is difficult to tell which fraction of this energy is emitted during the discharge as anomalous radiation or in the form of electromagnetic radiation of various frequencies, and which fraction is dissipated as heat. In general one should expect that some parts of the apparatus tend to heat up after repeated discharges. In practice, the massive copper constituents which are present, refrigerated by liquid helium, tend to stabilize the temperature. It was observed that if the discharges follow every two minutes (the charging time) the evaporation of helium is higher.

Let us summarize the features of the observed anomalous non-electromagnetic radiation (see Section 3).

1. It propagates in a well-collimated beam, with clean borders, having the same width as the superconducting emitter. The beam is emitted orthogonally to the electrode.

2. The radiation appears to propagate through brick walls and metal plates without noticeable absorption, but this is not due to a weak coupling with matter, because the radiation acts with significant strength on the test masses that are free to move. The electromagnetic shielding obtained with thick metal plates, the Faraday cage and UHF absorbing panels cannot be regarded as perfect; however, it is quite clear in our opinion that any residual electromagnetic radiation cannot be responsible for the observed impulse transfer to the targets (see also Point 3 below).

3. This radiation conveys an impulse which is certainly not related to the carried energy by the usual dispersion relation *E=cp*. One can in fact estimate, considering the data for the 18.5 grams pendulum, that the kinetic energy associated to the observed displacement is of the order of $10^{-4}$ *J* and the momentum is of the order of $10^{-3}$ *kg m/s*. If this momentum had to be imparted to the pendulum by radiation pressure, the energy needed in the beam would exceed the total energy available in the discharge (~$10^6$ *J* max). Moreover, the radiation energy in excess would heat up the pendulum (unless the pendulums are acting as perfect reflectors, which seems very unlikely).

4. The radiation acts upon targets of any composition, with a force proportional to their mass and apparently independent from their cross-section. The proportionality to mass is confirmed only within the reproducibility of the discharge process; the casual error (standard deviation of the single data) is between 5 and 7%. Further variations originate from inhomogeneities in the beam due to structural defects in the emitter (up to 12%). This implies that the force could be in fact not proportional to mass, but to some other quantity closely enough related to it, like for instance the baryonic number.

The choice of simple pendulums as detectors may appear rough and old-fashioned, but it does offer several advantages. Results with pendulums are straightforward and unambiguous, and precise enough in our case, given the casual variations already present in the beam intensity. Moreover, the use of pendulums allowed direct testing of the independence of the anomalous force on the composition of the targets. This was not possible with the microphone (or with piezoelectric detectors which can be used if measurements at higher frequencies are needed). Every precaution was taken to reduce any influence on the pendulums by acoustical or mechanical vibrations generated by the discharge much below the magnitude order of the observed effect. The propagation of the effect to a distance of 150 $m$ without measurable attenuation allows to rule out confidently all such spurious causes.

In order to estimate the strength of the beam's effect, suppose that $l$ is the length of a detection pendulum and $g$ is the local gravitational acceleration. Let $d$ be the half-amplitude of the oscillation. Let $t$ be the duration of the pulse, and $a$ the average acceleration imparted to the targets. $a$ has dimensions $m/s^2$. One easily computes that the product of the strength of the impulse by its duration is $at=\sqrt{(2gl(1-\sqrt{(1-(d/l)^2)}))}$. If $d<<l$, this formula can be simplified, and we have approximately $at\sim d\sqrt{(g/l)}d=2\pi d/T$, where $T$ is the period of the pendulum. With the data of Fig. 3, taking $t=10^{-4}$ $s$, one finds $a\sim 10^4$ $m/s^2$.

As remarked above (Point 3), the "dispersion relation" ratio $E/p$ of the observed radiation is very unusual. The term "radiation" is actually unsuitable, and one could possibly envisage an unknown quasi-static force field. In this way one could explain why an impulse is transmitted to the test masses. It is hard, however, to understand how such a field could be so well focused. This configuration cannot satisfy any known static field equation.

**4.2 Possible theoretical explanations**

A possible "gravitational" interpretation of the observed radiation is suggested by the proportionality of the force to the target mass, but is in strong contrast with the usual relationship between any gravitational field and its source. According to Einstein equations and General Relativity, only the matter-energy content of the source determines the gravitational field generated by it, not its particular state (like, for instance, presence or absence of superconductivity). In our case, even in the presence of vigorous discharges, the energy-momentum content of the apparatus is far too small, from the gravitational point of view, to generate observable effects. The repulsive character of the force is not explainable in the classical gravitational theory, either.

In a recent theoretical work [15] G. Fontana hypothesized that SN junctions involving cuprates could under certain conditions emit gravitational radiation. This is possible, in principle, because the pairs in cuprate superconductors are believed to be bound in a d-state, with angular momentum 2, and gravitons also have spin 2. The emission amplitude is however very small if computed through standard perturbation theory. Furthermore, it should be recalled that gravitational radiation is quadrupolar and cannot exert any net force on a target. The strength and the features of the observed radiation also seem to exclude any explanation based on gravitomagnetism or torsion in standard form [16], although some non-orthodox proposals in the literature suggest that these phenomena could be more relevant than generally believed [17].

An interpretation of the observed radiation in terms of gravitational fields leads to a conceptual difficulty if very massive targets are considered. Consider the case of a very massive pendulum placed in the beam path. If the effect is gravitational, then the acceleration of the test pendulum should not depend on its mass. However, it is clear that in order to give this mass the same oscillation amplitude of the small masses employed in the experiment, a huge energy amount is necessary, which cannot be provided by the device. Therefore the effect seems to violate the equivalence principle. Considering the back-reaction is probably necessary, namely the fact that the test mass exerts a reaction on the source of the impulse. This reaction is negligible as long as we use small test masses. Actually, the problem here is the inadequacy of the field concept in the presence of back-reaction – a problem that to our knowledge is usually absent in gravitation, due to the large size and mass of any relevant field source.

Let us now turn to the interpretation of the results obtained by observing the interaction between the anomalous radiation and a laser beam. If the anomalous radiation was electromagnetic, then according to classical and quantum electrodynamics it could not interact with light as observed, because the non-linear photon/photon coupling is exceedingly small in vacuum or in air. If the anomalous radiation was made of charged particles, then of course there would be a significant interaction with light; this hypothesis, however, is clearly ruled out by the rest of the observed phenomenology.

If we try to interpret the anomalous radiation in terms of a classical gravitational field, we run into contradiction with the standard general relativity theory. The field strength required to model the observed force on the pendulums is about $10^3 g$ for $10^{-4}$ s (see Section 4.1). In general, the effect of a classical gravitational field on light is of two kinds:

(1) Light is red- or blue-shifted. For instance, in the classical experiment by Pound and Rebka the upward propagation of γ-rays in the field of the Earth over a distance $h$ of 22.5 $m$ gave a relative red-shift $\Delta\lambda/\lambda \propto gh \sim 2.5 \cdot 10^{-15}$. In our case we would have $\Delta\lambda/\lambda \sim 5 \cdot 10^{-12}$, still too small to be detected by ordinary instrumentation.

(2) Light is bent by field curvature. If the anomalous radiation beam was a gravitational field, its curvature (of the order of the second space derivative of the field strength) would be located essentially at the borders. This would give a deviation angle θ of the order of $10^{-6}$ $rad$, again too small for detection (in general, the magnitude order of θ is θ~$d\sqrt{R}$, where $d$ is the size of the traversed curved region and $R$ is the curvature).

Finally suppose (compare Section 4.3 and the Conclusions Section) that the anomalous radiation consists of virtual gravitons with ν~$10^7$ $Hz$ and λ~$10^{-10}$ $m$. When interacting with the photons in a scattering process, such gravitons could easily expel them from the laser beam, because the momentum of optical photons is much smaller: $p=2\pi\hbar\lambda^{-1}$, with λ~$10^{-6}$ $m$.

**4.3 An anomalous vacuum effect?**

In view of the discussion above, and in the absence of any viable conventional explanations, we concluded that the observed radiation could be an anomalous vacuum effect. An effect, therefore, of quantum origin, and for some reason abnormally strong. We envisage an analogous of the Casimir effect, where small forces between uncharged metal plates are observed, which are due to a local inohomogeneity of electromagnetic vacuum fluctuations. There would be, of course, important differences with respect to the Casimir effect: (1) not electromagnetic, but gravitational fluctuations would be involved, since in our experiment the forces appear to be independent from the targets composition; (2) the situation is not static, but dynamic, with short transients involved; (3) the strength of the effect is abnormally large, and its range extraordinarily long.

It was shown by one of us in a series of theoretical papers [18] that the pairs condensate in superconductors can generate under certain conditions strong anomalous gravitational "dipolar" quantum fluctuations. In the present case, the critical conditions necessary for the anomalous coupling between the gravitational field and the pairs are produced only for a short time at the discharge. Therefore, the formation of virtual mass-energy dipoles could be accompanied by the emission of virtual dipolar radiation with frequencies of the same order of the frequencies present in the discharge. A beam of virtual gravitons, whose momentum originates from that of the pairs of the discharge, could be emitted, the beam direction being defined by momentum conservation. When hitting a target these gravitons could give rise to the observed impulsive forces.

In order to explain the features of the observed anomalous radiation and its propagation, one therefore could resort to the use of the "virtual radiation" concept. In quantum theory, any force is explained at a fundamental level as due to the exchange of virtual mediating particles. Technically, the expression for the interaction potential energy as the result of an exchange of virtual particles has been known for a long time in electrodynamics and in the gauge theories of weak and strong interactions. The corresponding formula for gravitation was given in [19], where the static interaction potential of two gravitational sources is expressed as an integral of the graviton propagator over momenta and energies. It is also possible to prove that the main contribution to the integral comes from momenta and energies related by the equality $E<<cp$, which corresponds quite well to what has been observed in the experiment but differs from the energy-momentum relation for free massless particles which can propagate to infinite distance independently from the existence of a target. In this formula the integral over momenta and energies of the exchanged particles is multiplied by the product of the two masses. This means that the flow of exchanged virtual gravitons does not have an "absolute" intensity, but is instead proportional to the source mass and to the target mass at the same time. Such a picture clearly differs from what one imagines in the case of a flux of real particles emitted from a source and absorbed by a target, and could explain the anomalous features of the observed radiation.

This picture could also explain the repulsive character of the force. From the quantum point of view, what matters in a scattering between two particles mediated by a third virtual particle (the simplest example being the electron/proton scattering, mediated by a virtual photon) is the exchanged energy-momentum. This is so true, that the signs of the electric charges, in the example above, do not affect the differential cross section: An electron bounces against a proton exactly the same way a positron does. This describes an elementary process where one single virtual particle is exchanged in a very short time interval. The static force is the opposite limit; many virtual particles are exchanged, in a continuous way, and in the time limit from $-\infty$ to $+\infty$. The phenomenon underlying the observed anomalous radiation could represent a sort of intermediate case, because we are supposing that many gravitons are exchanged, but still in a short time and so with the features of a scattering.

There is also the problem of how to explain the coherence in the beam. It resembles the coherence of a free electron laser beam (compare the Conclusions Section), even though in this case the beam is probably only virtual in the sense that its existence is limited in time and it does not have an absolute intensity. An analogy with optical lasers is impossible because the resonating cavity is missing.

## 5. Conclusions

The occurrence of the observed flat discharges is related to the crystal structure of the superconducting cathode. Since the high-conductivity *ab* planes are parallel to the surface, the electric field at this surface is extremely uniform. During the discharges intense supercurrents flow through the cathode in the *c* crystal direction. The superconducting charge carriers "leap" by tunnelling through $10^3$ or more *ab* planes in the discharge time. Due to macroscopic quantization, their wave functions keep a coherent phase in the motion. The non-electromagnetic radiation emitted in this process seems in fact to be spatially coherent and its wavelength is probably connected to the lattice spacing of the YBCO crystals. Constructive interference appears to be possible only if the radiation wavelength is equal to the spacing of the *ab* planes, i.e. $\lambda \sim 10^{-10}$ *cm*. Since the tunnelling frequency is $\nu \sim 10^7 \ s^{-1}$, we also may expect that this is the frequency of the radiation. These values for $\lambda$ and $\nu$ satisfy well the "integrated" relation $E<<cp$ observed at the detectors. It should be stressed, however, that we were not able to measure $\lambda$ and $\nu$ directly yet.

For a phenomenological description of the state of the superconducting emitter during the discharge, it would be possible to apply the Ginzburg-Landau model. In our case the model must be suitably modified, in order to take into account the anisotropy of the ceramic material; the model does not depend, however, on the microscopic mechanism of superconductivity in the material. In this context, identifying the behaviour of the superconducting charge carriers near the two surfaces which delimit the superconducting part of the emitter would be necessary. These surfaces are in contact from one side with the ionized gas during the discharge, and from the other with the ceramic not-superconducting layer, which has the same crystal structure as YBCO but is doped with a rare earth. Neither "boundary condition" is included in standard treatments available in the literature.

The propagation velocity of the radiation is still unknown, too. This can be measured in principle by placing two identical detectors A and B along the beam, at a known distance from each other (for instance, the maximum observed distance, AB=150 *m*). If the beam propagates with the speed of light, then the detection delay will be of the order of $10^{-6}$ *s*. This can be observed by comparing the signals of the two detectors as seen at the middle point between A and B. Then for a check one can exchange A with B. The method requires that the detectors have a temporal resolution better than $10^{-6}$ *s*.

In general, it is difficult to obtain fast rise times in detectors based on mechanical transducers. We recently started investigating another possibility, namely the interaction of the observed anomalous radiation with an optical laser beam. Some preliminary observations show that a measurable interaction actually occurs. Besides giving useful information about the nature of the anomalous radiation, this can allow its detection with fast-switching opto-electronic detectors applied to the laser beam after its interaction with the radiation. In this way we also expect to obtain in the future more precise information about the timing between the radiation emission and the phases of the discharge.

**Acknowledgment**

We are grateful to G. Fontana, C.Y. Taylor and H. Ucar for useful discussions.

# Figure Captions

**Fig. 1** - Discharge chamber (top view). The diameter of the chamber is about 1 *m*.

**Fig. 2** - General scheme of an Arkadjev-Marx high-voltage pulse generator. In our case the polarity is inverted.

**Fig. 3** - Correlation between the voltage discharge and the horizontal deflection of the pendulum.

**Fig. 4** - Impulse recorded by the microphone at a 67 degrees impact angle (amplitude in arbitrary relative units vs. time). Time scale is sampling period at $f_s = 44.1$ *kHz*. There is a 50 *Hz* noise due to power grid. The signal is unfiltered.

**Fig. 5** - Relative impulse strength versus impact angle.

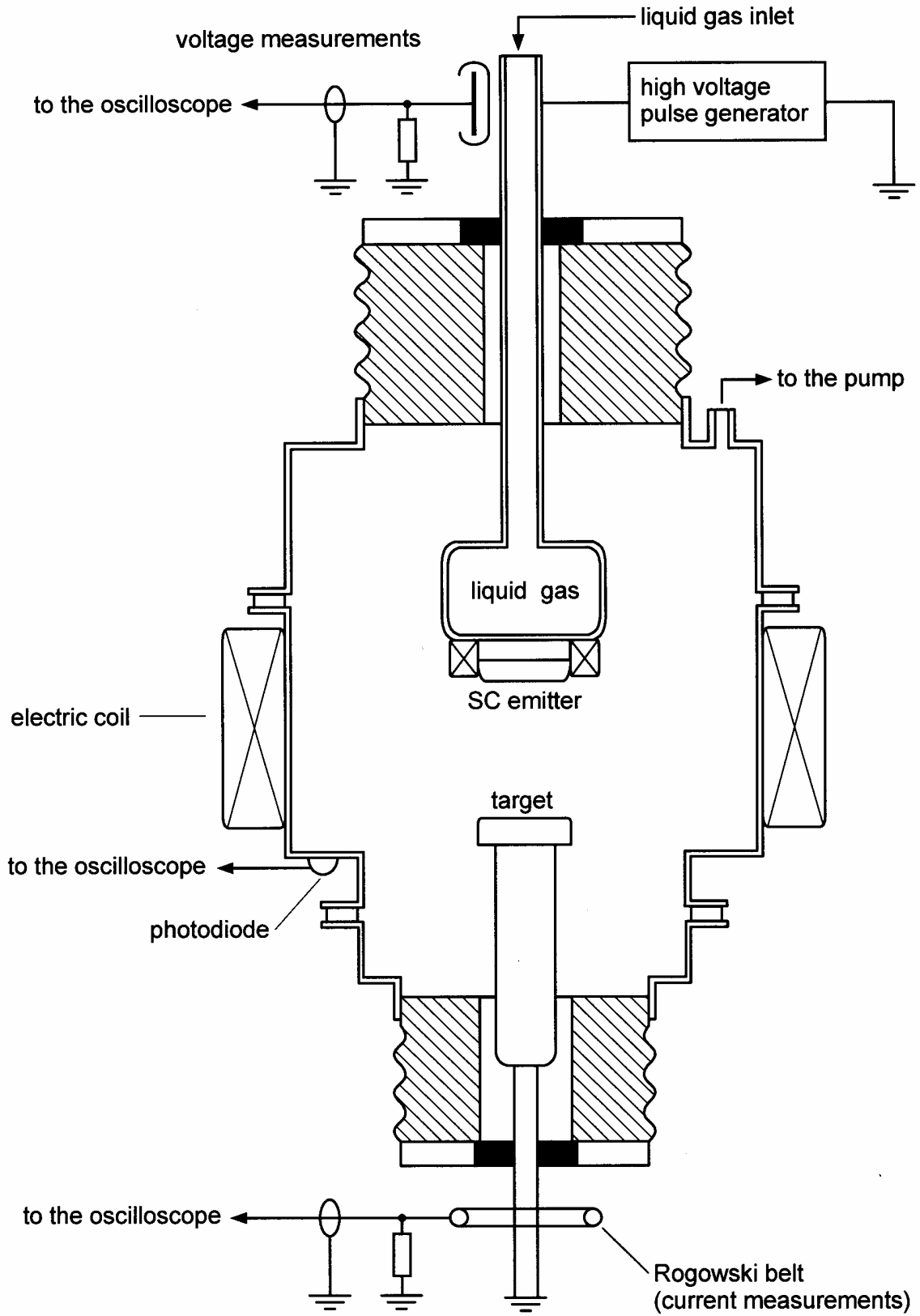

**Fig. 1**

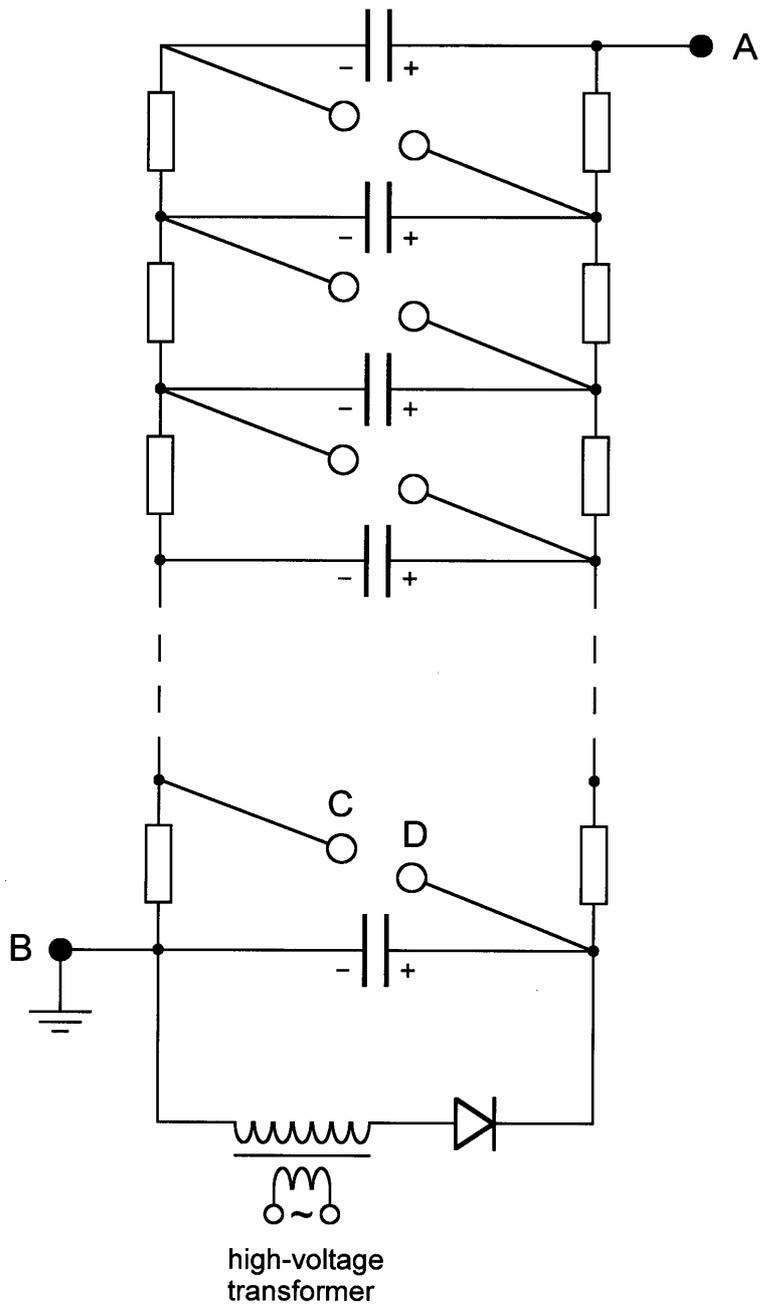

Fig. 2

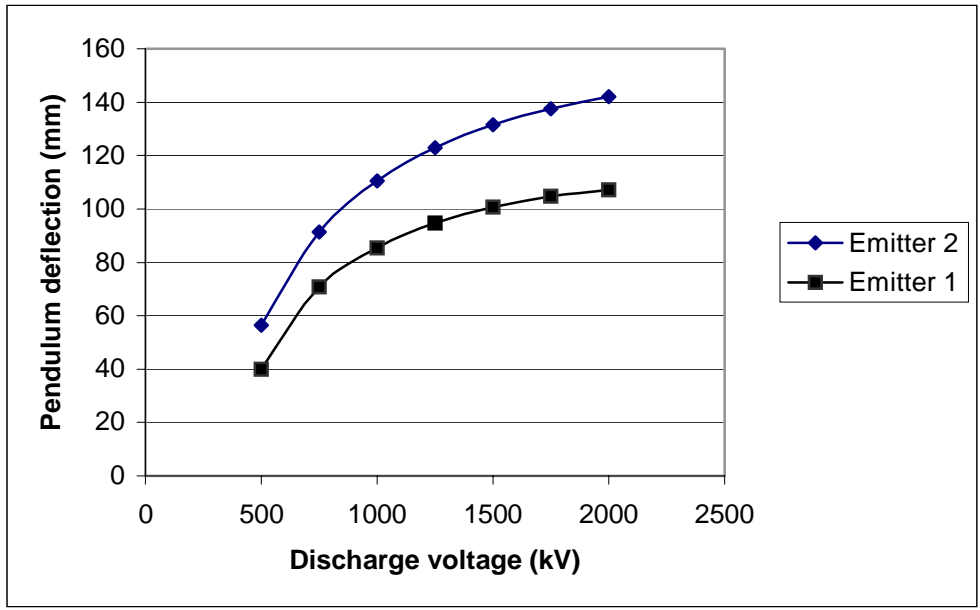

**Fig. 3**

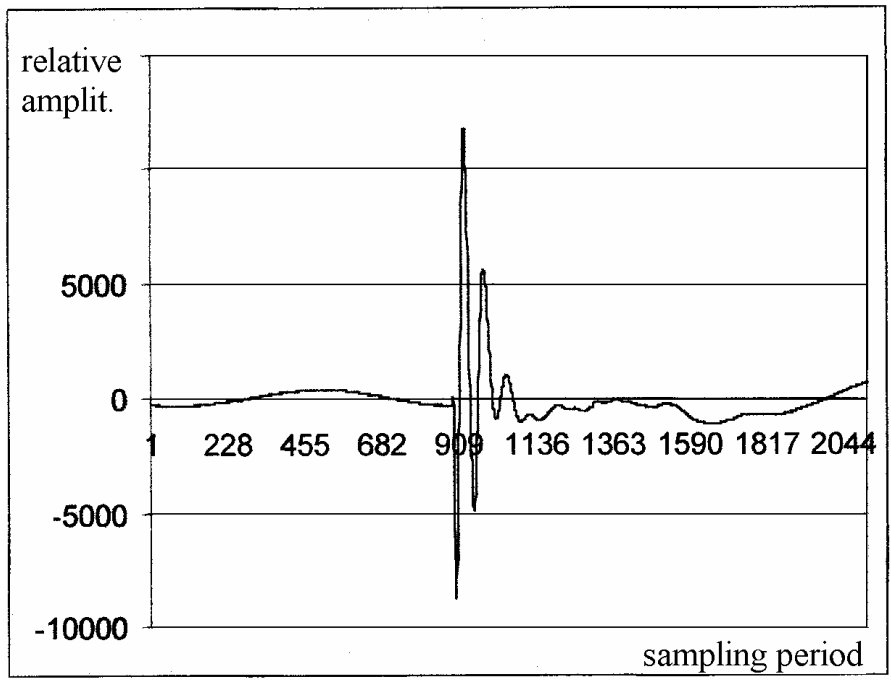

**Fig. 4**

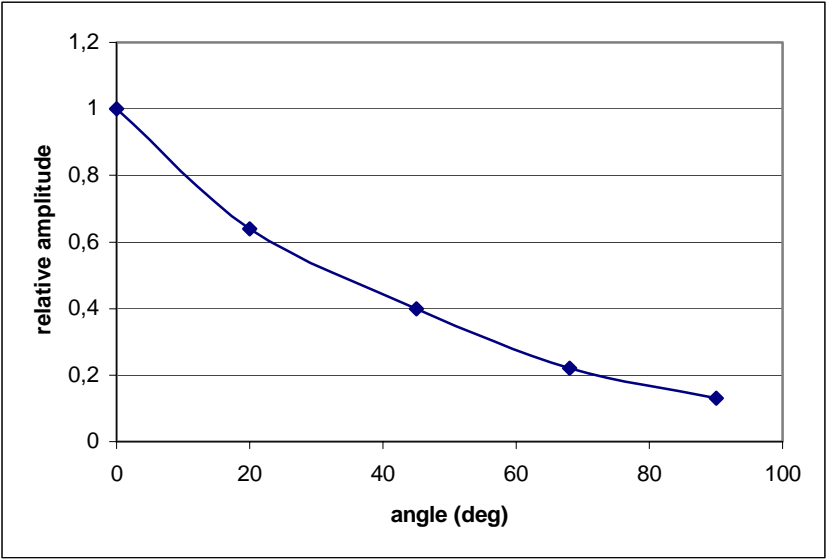

Fig. 5